\documentclass[doublecol]{epl2}
\usepackage{morefloats}
\usepackage{amsmath}
\usepackage{amsfonts}
\usepackage{graphicx}
\usepackage{dcolumn}
\usepackage{bm}
\usepackage{color}

\title{Emergence of (bi)multi-partiteness in networks having inhibitory and excitatory couplings
}
\author{Sarika Jalan \footnote{Corresponding author:sarikajalan9@gmail.com} and Sanjiv K. Dwivedi}
 \institute{Complex Systems Lab, Physics Discipline, Indian Institute of Technology Indore, Khandwa Road, Indore 452017, India}


\abstract{
(Bi)multi-partite interaction patterns are commonly 
observed in real world systems
which have inhibitory and excitatory couplings.
We hypothesize these structural interaction pattern to be stable 
and
naturally arising in the course of evolution. 
We demonstrate that a random structure evolves to the
(bi)multi-partite structure by imposing
stability criterion through minimization of the largest eigenvalue in the
genetic algorithm devised on the interacting units having inhibitory and excitatory couplings.
The evolved interaction patterns are robust against changes in the initial 
network architecture as well as fluctuations in the interaction weights.
}
\begin{document}
\maketitle

{\bf Introduction:} Finding mechanisms governing the evolution of patterns in real world systems remains challenging in 
the evolutionary science \cite{Levin}. Structural pattern of a network not only 
influences the functional response 
\cite{Boccaletti}, but also is motivated from a specific function of the corresponding system \cite{community}. 
Exploring evolutionary origins of existing structures is important to
develop an understanding pertaining to the hidden roles of these structures.
For example, nestedness in ecological networks structure has been shown as an evolved property in 
co-operative interactions \cite{Suweis} and also in maximizing the structural stability of mutualistic systems \cite{Rohr,Okuyama}. 

This Letter focuses on understanding the evolutionary origin of
bipartiteness and multipartiteness in real world networks. A network is 
called bipartite (multipartite) if its nodes can be divided
into two (more than two) groups in such a manner that nodes in one group are connected to the
nodes in the other groups with no or sparse connection existing within the same group. 
There are many natural systems which posses the bipartite (or multipartite) network structures. Few examples
of the real world networks, which are build
up to the current state as a result of evolution and possess
the bipartite (or multipartite) structure,
are gene regulatory interaction networks consisting of two groups 
of regulating and regulated
genes \cite{Milo}, ecological food webs having different trophic levels
\cite{Martinez,Mougi}, EI Verde rainforest \cite{Reagan}, etc.
However, understanding behind the organization of different trophic levels is
still not adequate \cite{Ingram}.

Additionally, several systems are known to be characterized by their constituents defined by
different behaviors.
For example, according to Dale's principle, in a neural system all outgoing synapses of a neuron 
are either inhibitory (I) or excitatory (E) \cite{Eccles}, thereby leading to various
structural properties due to I-E, I-I and E-E couplings \cite{Lazar,Economo}. 
Whereas, in ecosystem, several species show bias for the mixed behavior.
For example, Apex predators have zero or less number of predators for their own community \cite{Fabrizio}.
These
different behaviors of nodes or interactions are crucial for a desired
functioning or the robust functional performance of the underlying system \cite{Stefano,Allesina}.
The celebrated work by Robert May demonstrates that
the largest real part of eigenvalues ($R_{\mathrm{max}}$) of the corresponding Jacobian matrices
contain information about the stability of the underlying systems \cite{MayNature1972}. 
The emergence of complex structural properties 
has been studied using $R_{max}$ which is considered as a measure defining fitness
of the underlying system during evolution \cite{Perotti}.
Further, the Genetic algorithm (GA) is a widely used technique in optimization problems as well as in
providing evolutionary models for systems
in many disciplines \cite{Holland,GA_forrest}. It has been shown that a stability maximization-based
genetic algorithm leads to an emergence of the hierarchical modularity in a network \cite{Variano}.
Recently, evolution of clustering is demonstrated by maximizing the stability 
of the underlying network using GA
\cite{Jalan_genetic1}.
This paper provides an explanation to the 
emergence of bipartite (multipartite) structure using GA-based minimization of $R_{\mathrm{max}}$.
\begin{figure}[t]
\centerline{\includegraphics[height=3.5cm, width=5cm]{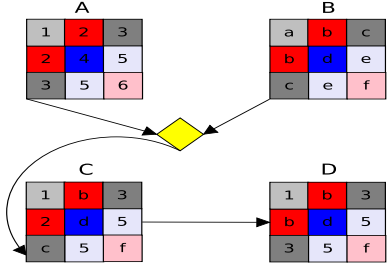}}
\caption{(Color online) Mechanism to generate a child
matrix from a pair of randomly selected fitter networks. The blocks of the parent and child matrices at a particular position are represented with the same color. The dimension of each square block is 10. Note that the GA works fine for all the block sizes significantly less than $N$. For larger block size, GA becomes meaningless as the child matrices lose variability and turn out to be very similar to their parent matrices.}
\label{Cross}
\end{figure}

{\bf Theoretical Framework:} We start with the $P$ independently drawn adjacency matrices, $[a_{\mathrm {ij}}]$ (each having dimension $N$) 
corresponding to Erd\"os-R\'enyi (ER) random networks representing the initial population. Elements in the corresponding adjacency matrices [$a_{ij}$] take value 1 and 0 depending upon whether there exists a connection between $ith$ and $jth$ nodes which is decided with a probability $q$.
\begin{figure}[h]
\centerline{\includegraphics[height=4cm, width=5cm]{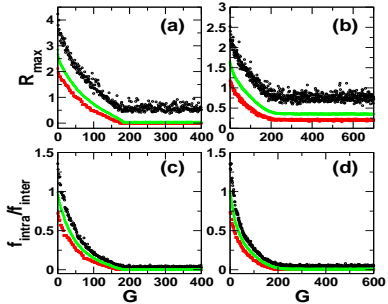}}
\caption{(Color online)
(a) and (c) evolution of
$R_{\mathrm max}$ and $f_{intra}/f_{inter}$ respectively for $p_{\mathrm{in}} = 0.50$ and
connection weight from Eq.~(\ref{rec_mat1}). Minimum, average and  maximum
values for the network population are indicated by $\square$ (red), $\diamond$ (green) and
$\circ$ (black)
respectively. (b) and (d) depict evolution
of $R_{\mathrm max}$ and $f_{intra}/f_{inter}$, respectively
for connection weight with fluctuations as imposed by
Eq.~(\ref{rec_mat2}).
Initial ER networks have $\langle k\rangle = 6$ with $N=100$. For all the cases network population is 500.}
\label{FigBipaER}
\end{figure}
Next, we generate matrices in which each node is either excitatory (type-I) or inhibitory (type-II) throughout the evolution.
This is motivated from the Jacobian (synaptic) matrices considered in the model of \cite{RajanAbott}, which incorporates Dale's principle. Such type of matrices have following properties: (i) The columns are either Gaussian distributed numbers with negative mean or positive mean with certain variance. (ii) Summation of the numbers in each row is equal to zero, which takes into account the balanced condition. Such matrices lead to the underlying networks that are globally connected.
Therefore, with probability $p_{\mathrm {in}}$, type-II nodes (i.e. nodes with only I connections) are randomly selected out of
the total $N$ nodes (Eq.~\ref{rec_mat1}). Rest of the nodes form a set of the type-I nodes (i.e. nodes with only E connections) of the behavior matrices.
This arrangement leads to the following matrix, termed as behavior matrix, which has a balance of the
inhibition and excitation \cite{partII,RajanAbott} in the whole network,
\begin{equation}
b_{\mathrm{ij}} = \begin{cases}\begin{aligned}1-1/p_{\mathrm{in}} \forall j, \mbox{~~if~i~$\in$~type-II}\\
 1~~~~~~~~~~ \forall j, \mbox{~~if~i~$\in$~type-I}\end{aligned}\\
\end{cases}
\label{rec_mat1}
\end{equation} 
More specifically in Eq.~\ref{rec_mat1}, rows of matrices [$b_{\mathrm{ij}}$] have either `+1' or 
`$1-1/p_{\mathrm {in}}$' entries.
The entries of the $b_{\mathrm{ij}}$ matrices are fixed in the course of evolution, demonstrating the 
invariant nature of nodes and consequently the links \cite{Schaffer}.
We define another matrix [$c_{\mathrm {ij}}$] (Eq.~\ref{rec_mat2}) for assessing the fitness of a network, which is constructed using 
the adjacency matrices of the network and the behavior matrix (Eq.~\ref{rec_mat1}) in the following manner:
\begin{equation}
c_{\mathrm {ij}} = \begin{cases} b_{\mathrm {ij}}~~\mbox{if } a_{\mathrm {ij}}\not=0 \\
0 ~~ \mbox{otherwise}  \end{cases}
\label{rec_mat2}
\end{equation}
\begin{figure}[t]
\centerline{\includegraphics[height=2cm, width=5cm]{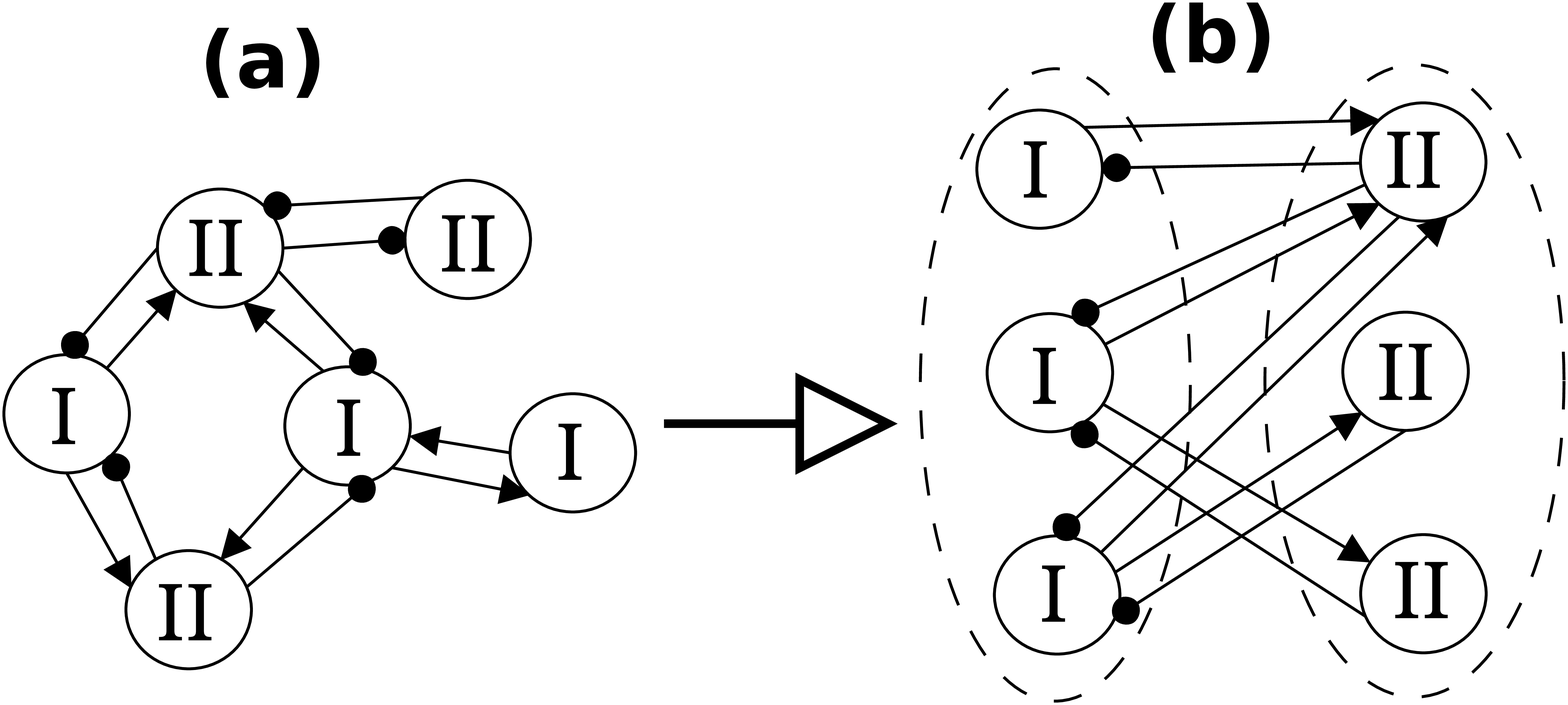}}
\caption{Schematic diagram for (a) initial population of ER networks and
(b) finally evolved bipartite structure (dashed circles).
Dots and arrows represent
I and E links, respectively.}
\label{Random_Bipa}
\end{figure}

A network corresponding to lower $R_{\mathrm{max}}$ value of an associated [$c_{\mathrm {ij}}$] matrix is referred here as a fitter network.
For the next generation, the $P$ fitter networks of size $N \times N$ each are created as follows.
First of all we select $P/2$ number of distinct networks from these networks which
have $R_{\mathrm {max}}$ values
lesser than those of the rest of the networks in the present population. These selected networks are 
fitter networks and act as the parent networks for creating rest $P/2$ child
networks for the next generation (Fig.~\ref{Cross}).
For example, in order to generate one child (say C), a pair
of networks (say A and B) from $P/2$ population are selected randomly. Then each block of the adjacency matrix of the child at a specified 
dimension and location (in terms of rows and 
columns), is filled with the block having the same
position in the parent matrices, with an equal probability (Fig.~\ref{Cross}).
The child matrices generated are asymmetric in nature. We thus take the upper triangular part of the child matrices and construct their adjacency matrix such that they are symmetric. In order to avoid trivial changes in the $R_{\mathrm{max}}$ values due to the change in the total number of connections in the network and to get a signature of the change in the network pattern on the $R_{\mathrm{max}}$ values, we preserve the average degree ($\langle k \rangle$) of the child matrices by randomly removing or inserting connections with a small probability. This probability is decided by the fluctuation of the average degree in the crossed child network from the degree of the initial networks population used in the GA. Thus, the total number of edges remain almost same for the child networks.

In order to emphasize that here changes in the value of $R_{\mathrm{max}}$ are
not due to the change in the average degree \cite{book_spectra} but arising due to the changes
in the interaction patterns, we present the results for
the preserved average degree of the child network. 
Since real networks are known to
be sparse in nature, the results are presented for $\langle k \rangle = 6$. A lesser value of
$\langle k \rangle$ leads to disconnected components, in turn affecting $R_{\mathrm{max}}$.
With an increase in
$\langle k \rangle$, $R_{\mathrm{max}}$ increases in general.
The larger $\langle k \rangle$ leads to a less variation among the structural conformations of initial network population, which results in a lesser variation in the child population as well. The small amount of variation present in the set of initial networks population vanishes after certain generations without producing sufficient structural changes with respect to the initial population.
The evolution stops when there is no structural variation in the network population.
Also bipartite structure does not exist for a network having $\langle k \rangle$ $> N/2$. Thus, the GA approach which preserves the average degree, does not lead to evolution of bipartite structure for very dense networks.
Further, with larger $N$, the values of $R_{\mathrm{max}}$ may
vary but the overall behavior that $R_{\mathrm{max}}$ minimization leads to evolution of the bipartiteness is expected
to remain same, provided the initial networks consists of different types of nodes. The problem encountered on considering larger network size \cite{Variano},
would be requirement of memory allocation for a population of $P$ networks in GA amounting to
$PN^{2}$ size of the 1-d array, which is computationally not favorable for large $N$.

Further, the total number of connections within the 
same population (within type-I or within type-II) is denoted
by $f_{\mathrm {intra}}$ and
the total number of connections among the different populations (between type-I and type-II) is denoted by $f_{\mathrm {inter}}$. The ratio of $f_{\mathrm {intra}}$ and $f_{\mathrm {inter}}$ provides a measure for the bipartivity \cite{Estrada,Holme}.
\begin{figure}[h]
\begin{center}
\centerline{\includegraphics[height=2.5cm, width=4cm]{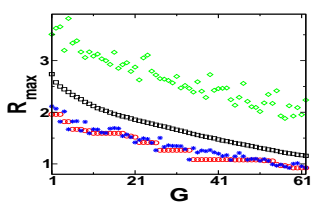}}
\caption{(Color online) $R_{\mathrm {max}}$ during the evolution of matrices
[$c_{\mathrm {ij}}$] for children and parents population.
Square (black) and circle (red) represent maxima and minima of
$R_{\mathrm {max}}$ values of the [$c_{\mathrm {ij}}$] matrices
for the parents population respectively and
diamond (green) and star (blue) are maxima and minima of
$R_{\mathrm {max}}$ values of [$c_{\mathrm {ij}}$] matrices
for the children population respectively. }
\label{Par_Child}
\end{center}
\end{figure} 
The zero value of
$f_{intra}/f_{inter}$ 
indicates a bipartite structure.
However the converse is not true, a bipartite structure might also arise when in a network, there are
two groups of nodes such that one group (having both type-I and type-II category nodes) forms all the connections with the other group (again having both type-I and type-II category nodes) and no connection within the group. Connections between type-I and type-II nodes of individual groups would contribute to
$f_{intra}$ leading to non-zero $f_{intra}/f_{inter}$ value. Note that,
a bipartite structure may exist in a system having only single behavior. 
For example, a system in which plants and pollinators form two different communities exhibiting the bipartite structure with only type-I couplings 
\cite{Suweis,Rohr,Okuyama,Allesina}. This situation does not arise in our analysis as we consider different classes of nodes and bipartiteness is confirmed only when the value of $f_{intra}/f_{inter}$ is zero.

{\bf Emergence of bipartite structure:} 
Starting with sparse ER random networks as an initial population, as evolution progresses through GA (Fig.~\ref{FigBipaER}), the minimization of $R_{\mathrm{max}}$ occurs.
In the initial population, the mean of $f_{intra}/f_{inter}$ is close to $one$, however maxima and minima 
are different as expected from the various realizations of the ER networks. As evolution progresses, the mean and minimum values of 
$f_{intra}/f_{inter}$ converge towards zero, however
maxima still exhibits a small separation from the zero (Fig.~\ref{FigBipaER}). This indicates that
most of the networks in the evolved population attain the bipartite structure
in which no or a very few connections exist within type-I or within type-II category nodes. Since, the GA minimizes $R_{\mathrm{max}}$ values of the population, 
the convergence of the maxima towards zero is stopped when the minima closely 
coincides with
the mean value. What follows that the initial population consisting of random architecture (Fig.~\ref{Random_Bipa}(a)) 
converges to the bipartite structure (Fig.~\ref{Random_Bipa}(b)) through the stability maximization.

The emergence of the bipartite structure directly follows from the
 spectral properties of antisymmetric matrices entailing all imaginary eigenvalues.
The [$c_{\mathrm {ij}}$] matrix, for a ideal bipartite structure without any connection within type-I or within type-II category nodes, corresponds to a antisymmetric matrix, i.e. [$c_{\mathrm {ij}}$ = - $c_{\mathrm {ji}}$]. 
The $R_{\mathrm{max}}$ of [$c_{\mathrm {ij}}$] for the initially considered ER 
networks, generated using connection probability $q$, takes positive values given by $\sqrt{Nq}$ for $p_{\mathrm{in}}$ = 0.50 \cite{pre2011}. 
Additionally, the correlation in the elements of $c$ matrix, given as 
$\tau = \sum_{i,j=1}^{N} c_{\mathrm{ij}}c_{\mathrm{ji}}/\sum_{i,j=1}^{N} c_{\mathrm{ij}}c_{\mathrm{ij}}$ for the matrices defined by Eq.~\ref{rec_mat1}, takes value 
\begin{equation}
\tau = (f_{\mathrm {intra}} - f_{\mathrm {inter}})/(f_{\mathrm {intra}} + f_{\mathrm {inter}})
\end{equation}
for $p_{in}=0.5$. This value decreases with $f_{intra}/f_{inter}$, 
further leading to a decrease in $R_{\mathrm {max}}$ because the spectral 
distribution of [$c_{\mathrm{ji}}$] matrices are of elliptical shape with its axis lying
on the real line decreasing around a fixed center \cite{Sompolinsky}.
Since we minimize  $R_{\mathrm {max}}$, the GA successfully and smoothly evolves a structure in 
which $f_{intra}/f_{inter}$ has very small or zero value.

The successful performance of the GA can be explained by considering two different classes of 
network population.
Let the children and their parent populations form two different classes of networks in each generation and the $R_{\mathrm {max}}$ of [$c_{\mathrm {ij}}$] associated with both the classes, decrease during the evolution.
The maxima and minima of $R_{\mathrm {max}}$ for the parent [$c_{\mathrm {ij}}$] 
decrease as per the GA minimization (Fig.~\ref{Par_Child}).
The maxima and minima of $R_{\mathrm {max}}$ for children [$c_{\mathrm {ij}}$] also decrease with 
the evolution. It happens so that the maxima of 
$R_{\mathrm {max}}$ values for children population is greater  
than the maxima of those for the parents population. 
The hamming distance between the adjacency matrices of networks provide a method to measure this 
dissimilarity \cite{Tun}. What follows that, whenever parents exhibit
dissimilarity, i.e. Hamming distance is large, their crossed child is also likely to have a
high deviation in its structural properties which in turn leads to a large deviation 
of $R_{\mathrm {max}}$ of the child from its parent.

Furthermore, the minima value for children and parents populations may lead to the following
two cases.  
In the first case,
the minima value for children is higher than the minima value for the parent population. 
Hence, the minima value for parent population in the subsequent generations
does not decrease until the minima values for the children become lower than the minima value of their
parents population. Whereas in the second case, the minima values for parents are higher
than those of child, hence parents, which produce child for the 
next generation, are selected from the child in the current generation leading to a decrease
in the minima of the parents for the next generation.
This leads to the value of $R_{\mathrm {max}}$ decreasing in 
each generation which indicates that as evolution progresses, 
the evolved feature (i.e. a lesser value of $f_{intra}/f_{inter}$)
is transferred to the child. It occurs due to the procedure of block selection
for child matrices as
represented in the schematic diagram (Fig.~\ref{Cross}). 
The GA-based optimization method converts the entries of the adjacency matrices corresponding to the connections within type-I nodes and the connections within type-II nodes to E-I couplings, thus leading to a decrease in $f_{intra}/f_{inter}$ value.
The decreased values of $f_{intra}/f_{inter}$ of parent matrices give rise to a chance to produce a child
with a further decreased value. And whenever parents produce such type of child network, they
possess lower $R_{\mathrm {max}}$ values which are selected for the next generation.
In other words, fitter children are produced by the fitter parents.

{\bf Robustness against random fluctuations:} To demonstrate the robustness of the results against changes in coupling strength, we introduce
random fluctuations \cite{Allesina,Yuan}, which can be imposed in the fitness matrix as follows
\begin{equation}
c_{\mathrm {ij}} = \begin{cases} Xb_{\mathrm {ij}}~~\mbox{if } a_{\mathrm {ij}}\not=0 \\
0 ~~ \mbox{otherwise} ~~a_{\mathrm {ij}}=0. \end{cases}
\end{equation}
where $X$ is a uniform random variable lying between 0 and 1.
As illustrated in Fig.~\ref{FigBipaER}, the bipartite structure evolved through
evolution remains unaffected by the random 
fluctuation in coupling strength. The only implication of having the random fluctuations
is that the evolutionary process becomes bit 
slower.
The minima, maxima and mean of $R_{\mathrm{max}}$ 
decrease during the course of evolution with the mean being bit closer to the minima rather than the maxima (Fig.~\ref{FigBipaER}).
After few generations, three values become saturated i.e. parallel to
the generation axis. The minima does not converge to zero which is unlikely for no 
fluctuations case. The structural changes during
the evolution lead to the decrease in the minima, mean and maxima, 
whereas $f_{intra}/f_{inter}$ still remains close to zero similar to the no fluctuation case.
\begin{figure}[t]
\centerline{\includegraphics[height=2.5cm, width=8cm]{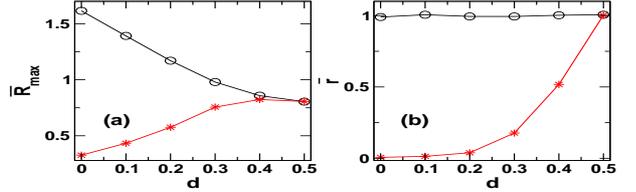}}
\caption{(Color online) (a) Average $R_{\mathrm {max}}$  and
(b) $r = (f_{intra}/f_{inter})$ for both the initial (circles),
and the evolved (stars) networks as a function of $d$.
The system is evolved for 5000 iterations consisting of 500 networks
population for GA.
The average values, for the evolved networks, is calculated 
for all the population of last 1000 generations.
Initial ER networks have $\langle k\rangle$ = 6 and $N=100$. }
\label{BothsideF}
\end{figure}
 
Further, in order to analyze the effect of fluctuations in the
coupling nature of the predefined coupling behaviour in 
matrix ([$b_{ij}$]), we take the 
range of uniform random variable in the negative side as well
using the following equation;
\begin{equation}
Y = X - d
\label{Randsift}
\end{equation}
where 
$d$ is a number lying between 0 and 0.5.
Therefore, fitness matrix [$c_{ij}$] would be defined as;
\begin{equation}
c_{\mathrm {ij}} = \begin{cases} Yb_{\mathrm {ij}}~~\mbox{if } a_{\mathrm {ij}}\not=0 \\
0 ~~ \mbox{otherwise} ~~a_{\mathrm {ij}}=0. \end{cases}
\label{rec_Randsift}
\end{equation}
We run the simulations for various $d$ values in which fitness matrix is defined by Eq.~\ref{rec_Randsift}.
For $d = 0.5$, 
matrices completely destroy the impact of predefined information of the 
behavior matrix([$b_{ij}$]) during the evolution from Eq.~\ref{Randsift}. What follows that
we do not get any structural
changes during the evolution and hence there is no decrease in $\overline{R}_{max}$. 
However, for $d < 0.5$, the impact of behavior matrix on [$c_{ij}$] matrix 
is dependent on the value of $d$.
Fig.~\ref{Randsift}(a) depicts that the efficiency of optimization
depicted by difference in the value of $\overline{R}_{max}$ of the evolved 
and the initial networks, increases  
with a decrease in the $d$ values. For $d = 0.5$, the rate of selection and 
rejection of the networks in a given generation
becomes equal and very fast due to equal strength of the fluctuations 
in the positive and negative side of the entries
of the fitness matrix [$c_{ij}$] for any predefined coupling behaviour
([$b_{ij}$]). Consequently, GA fails to minimize $\overline{R}_{max}$ value  and as
a result no variation in values of $f_{intra}/f_{inter}$ is observed at this
value of $d$
(Fig.~\ref{Randsift}(a)-(b)). As $d$ decreases,
$f_{intra}/f_{inter}$ decreases with a  
faster rate in the evolved networks (Fig.~\ref{Randsift}(a)). 
After a certain value of $d$ the rate of decrement becomes slower. Therefore,
there exists a large regime of $d$ in which the 
evolved networks exhibit the bipartite or close to the bipartite
structure (Fig. \ref{Randsift}(b)).
\begin{figure}[t]
\centerline{\includegraphics[height=4cm, width=5cm]{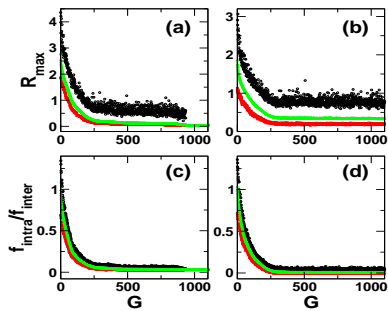}}
\caption{(Color online) Evolution of bipartite structure for system having
both types of the nodes, same as that of Fig.~\ref{FigBipaER} for the initial SF networks
with $\langle k \rangle =  6$ and $N=100$.
The network population used in GA is 500.
(a) and (c) present the evolution of $R_{\mathrm {max}}$ and $f_{intra}/f_{inter}$, respectively, for fitness matrix given
as Eq.~2, whereas (b) and (d) present the evolution of these quantities for fitness matrix being
represented by Eq.~4.}
\label{FigBipaSF}
\end{figure} 

In order to demonstrate the robustness of results against changes in the initial population networks,
we study the evolution of scale-free networks. These networks are generated using 
Albert-Barabasi model \cite{rev_barabasi} for both the cases of with and without fluctuations in the coupling strength.
Since the scale-free networks  have randomness arising due to the algorithm, it leads to 
the appearance of structural variations in different realizations.
The given position of blocks in associated
adjacency matrices have sufficient variations, as a result the crossed child matrices in GA attain conformation structure leading to bipartite arrangement.
For coupling
with random fluctuations, there is again an emergence of the bipartite structure (Fig.~\ref{FigBipaSF}). 

We have also analyzed the case when the directed networks form the initial population for the GA. For random directed networks, there would be a very less probability of finding un-directional ($A_{ij}$=$A_{ji}$) coupling and this probability is dependent on the size and average degree of the network. However, in real world networks, such un-directional couplings are found with very high probability \cite{Hagen_2012}.
For initial networks being directed, the evolved networks
exhibit bipartiteness as exhibited by undirected initial networks as long as
the child networks of each generation preserve directionally possessed by the
parent networks. The evolution to the bipartiteness as a result
of stability maximization directly follows from the discussions around Eq.~3.
The only 
difference is that it requires more number of generations to obtain the evolved networks as compared to the undirected case.
\begin{figure}[t]
\centerline{\includegraphics[height=2.5cm, width=8cm]{ECOmodified.eps}}
\centerline{\includegraphics[height=2.5cm, width=8cm]{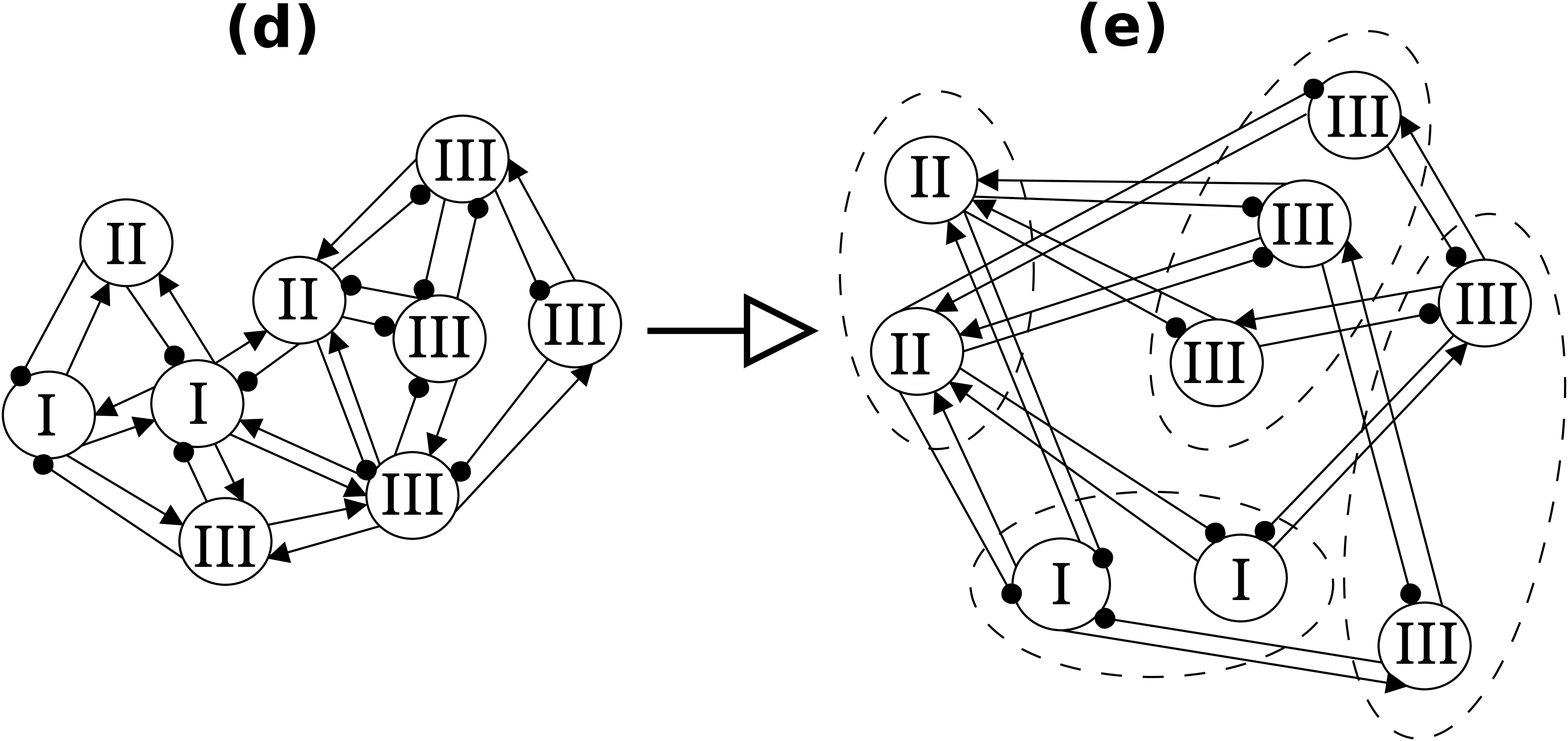}}
\caption{(Color online) (a) Average $r$
 for the evolved bipartite networks as a function of $p$. The population at each step consists of 500 networks.
(b) Change in the average value of total number of edges among type-I and type-II nodes over the network population depicted by $\overline{g}_{intra}$ with the evolution. Zero value of $g_{intra}$ indicates formation of different trophic levels consisting type-I and type-II nodes each.
(c) Average behavior of $\overline{R}_{\mathrm{max}}$ over network population with the evolution.
Calculations are done for the size of network being 100 and average degree 6
with type-I, type-II and type-III being 20, 20 and 60, respectively. Network population used for (b)-(c) is 1000. For all the cases, $[c_{ij}]$ matrices are defined by Eq. 4. (d) Initial population of ER networks when all three types of nodes (type-I, type-II and type-III) coexist (e) finally evolved structure consisting different trophic levels (dashed
circles).
 Dots and arrows represent
I and E links, respectively.}
\label{eco}
\end{figure}

{\bf Addition of type III nodes and emergence of multipartite trophic networks:}
The results presented above restrict nodes with only inhibitory and excitatory connections, which is motivated from neural models \cite{RajanAbott}. If we relax this criterion, nodes may have inhibitory as well as excitatory connections leading to characterization of nodes as type-III (Fig.~\ref{eco}(d)). 
We investigate the evolution of such type of networks.
 Starting with a matrix consisting of $1$ and $-1$ entries randomly distributed, which corresponds to all nodes being type-III, there may be some rows having more $1$ and some rows having more $-1$ entries. Now we select a row having more $1$ entries and with probability $p$ convert $-1$ entries into $1$ entries. Similarly, for rows having more $-1$ entries, with probability $p$ we convert $1$ entries into $-1$ entries. The value of $p$ being zero corresponds to the original matrix which we have started with denoting all type-III category nodes, whereas $p=1$ leads to the situation where those rows having more $1$ entries in the initial matrix, have all their $-1$ entries converted into 
$1$ corresponding to type-I nodes and those rows having more $-1$ entries in the initial matrix, have all their $1$ entries converted into $-1$ corresponding to type-II nodes. We find that with
increase in $p$, there is increase in bipartiteness in
the evolved networks (Fig.~\ref{eco}(a)).

In the above case, for all values of $p$ either we have only type-III nodes or type-I and type-II nodes with a feeble probability of co-existence of all three types. Whereas in a realistic situation, for instance in ecological systems, there exist few species which are only prey (type-I), few others which are only predator (type-II) and some more species which are prey to some species and act as predators for others (type-III), for example, the herbivores in the grassland food web \cite{Estrada}. In order to mimic this realistic situation which is a mixture of all three types, we consider some of the nodes as type-I, some of the nodes as type-II and rest as type-III. Again type-I nodes will lead to all $1$ entries and type-II nodes will lead to all $-1$ entries in the behavior matrix. In rows corresponding to type-III nodes, $1$ and $-1$ entries are introduced with the equal probability.

We find that with an increase in the generations, connections within type-I nodes and connections within type-II nodes decrease and converge towards zero (Fig.~\ref{eco}(b) and Fig.~\ref{eco}(e)).  
This leads to a situation when type-I nodes have type-II or type-III neighbors and type-II nodes have type-I or type-III neighbors. 
GA minimizes $R_{\mathrm{max}}$ with relatively faster rate in the beginning of the evolution and as
the evolution progresses the rate becomes slower (Fig.~\ref{eco}(c)) due to a steep decrease in the value of
$\tau$ \cite{Jalan_genetic1}.
Further, number of the type-III nodes, 
which are neighbors of the type-I nodes but are not neighbors of the type-II or vice versa, are maintained with random fluctuations in the course of evolution. Additionally, with the evolution,
the number of connections among such type-III nodes decreases.
The nodes of type-I and type-II are not connected within their categories 
and thus form two distinct groups
 (Fig.~\ref{eco}(e)).
Also, the type-III nodes, which are neighbors of type-I (type-II) nodes but are not the neighbors of type-II (type-I) nodes (Fig.~\ref{eco}(e)), form other group(s). This organization resembles the various trophic levels of food-web
where the species belonging to same trophic level do not have any interaction among themselves \cite{Martinez,Mougi}

{\bf Conclusion:}
To conclude, we present an evolutionary origin of the emergence of bipartiteness in 
interaction patterns. The essence of the method lies in the property that individuals get classified in to different
classes in which 
they have fixed invariant behaviors. The $R_{max}$ values are dependent on network architecture as well as fluctuations in coupling strength. Since $R_{max}$ quantifies the stability of a system \cite{MayNature1972}, there arises a chance of this value to increase, making a system unstable.
Additionally, an increase in the system size may also lead to an increase in $R_{\mathrm{max}}$ implying further that a system with larger size ideally should not exist \cite{MayNature1972}.
However, our results suggest that despite larger system size and high disorder in coupling strength, a balance of I and E limits $R_{\mathrm{max}}$ in the evolved networks, and as a result the fluctuations may not destabilize the system.  This result provides an insight in to why despite possessing  
high degree of disorder ecosystems are robust, which can be attributed to the existence of various trophic levels of species
preserving stability of the system.

The method considered here may be useful in optimizing various other measures
related with structural as well as spectral behavior \cite{Liu,Gang}.
The framework can be extended further to derive an evolutionary understanding of how a systems's function gets affected by underlying structural patterns.
For instance, assortative-disassortative mixing interaction  patterns are
seen in real world networks \cite{community},
evolution of which can be addressed from the stability point of view.

{\bf Acknowledgements:} 
SJ thanks DST  
(EMR/2014/00368) and CSIR (25(0205)/12/EMR-II) for funding.
SKD acknowledges the University grants commission, India for
financial support.

\end{document}